\documentclass[useAMS,usenatbib]{mn2e}
\usepackage{amssymb}
\usepackage{graphics}

\def\eqref#1{equation~(\ref{#1})}
\def\eqrefs#1#2{equations~(\ref{#1})-(\ref{#2})}
\def\apref#1{Appendix~\ref{#1}}

\def\binom#1#2{{#1 \choose #2}}
\def\det{\mathop{\mathrm{det}}\nolimits}
\def\eoq{\mathrm{q}}
\def\eop{\mathrm{p}}

\title[An analytical solution for Kepler's problem]{An analytical solution for Kepler's problem}
\author[A. P\'al]{%
Andr\'as P\'al\thanks{E-mail: apal@szofi.net}\\
Harvard-Smithsonian Center for Astrophysics,
	60 Garden street,
	Cambridge, MA, 02138, USA \\
Department of Astronomy, Lor\'and E\"otv\"os University,
	P\'azm\'any P. st. 1/A,
	Budapest H-1117, Hungary \\
Konkoly Observatory of the Hungarian Academy of Sciences, 
	Konkoly Thege Mikl\'os \'ut 15-17,
	H-1121 Budapest, Hungary}

\begin{document}

\date{Accepted \dots, received \dots; in original form \dots}

\pagerange{\pageref{firstpage}--\pageref{lastpage}} \pubyear{2007}

\maketitle

\label{firstpage}

\begin{abstract}
In this paper we present a framework which
provides an analytical (i.e., infinitely differentiable) transformation
between spatial coordinates and orbital elements for the 
solution of the gravitational two-body problem. The formalism omits
all singular variables which otherwise would yield discontinuities.
This method is based on two simple
real functions for which the derivative rules are only required to be
known, all other applications -- e.g., calculating the orbital
velocities, obtaining the partial derivatives of radial velocity
curves with respect to the orbital elements -- are thereafter straightforward. 
As it is shown, the presented formalism can be applied to find
optimal instants for radial velocity measurements in transiting
exoplanetary systems to constrain the orbital eccentricity as well 
as to detect secular variations in the eccentricity or in the 
longitude of periastron.
\end{abstract}



\section{Introduction}
\label{sec:introduction}

In recent years, precise radial velocity (RV) observation of stars 
and careful analysis of RV data have become relevant in 
astrophysics since the vast majority 
of the known extrasolar planets have been discovered using this method
\citep{mayor1995}
or have been confirmed this way when the planet itself was first
detected as a transiting object around its host 
star \citep{konacki2003}. 
In the cases where there are no observable transits,
the characterization of extrasolar planets relies on the
radial velocity observations alone\footnote{Of course, 
with the exception of planets discovered by microlensing or direct imaging.
The characterization scheme of planets around pulsars are highly similar
to the analysis of RV data.}.
Moreover, observations of photometric transits 
in addition to radial velocity measurements
constrain the mass of the planet (instead of yielding only a lower
limit), and provide more precise information on the 
epoch and period; see e.g. the case of the low-mass transiting planet
HAT-P-11b \citep{bakos2009}, where the uncertainty in the epoch
would be $\sim 500$ times larger if the analysis had relied only 
on radial velocity measurements.
Thus, incorporating constraints given by transit timings
reduce the uncertainties in the RV amplitude and the orbital 
parameters (e.g., semimajor axis, eccentricity).

The aim of this paper is to present a set of analytic relations (based 
on a few smooth functions defined in a closed form) 
which provides a straightforward solution of Kepler's 
problem, and consequently, time series
of RV data and RV model functions. Due to the analytic property, the
partial derivatives can also be obtained directly and therefore can be
utilized in various fitting and data analysis methods, including, for 
instance, the Fisher analysis of covariances, uncertainties and correlations.
The functions presented here are nearly as simple to manage 
as trigonometric functions. 

As an application, we give a detailed description about scheduling
radial velocity measurements in the case of transiting 
extrasolar planets in order to derive accurate orbital eccentricity
and/or detect the variations in the eccentricity. Discussions
related to this problem in the case of extrasolar
planets with no additional constraints on their orbital motions
other than RV data can be found in \cite{loredo2003},
\cite{ford2008} or \cite{baluev2008}.
Photometric measurements for transits constrain the orbital period
and epoch much more precisely than pure RV observations. Thus,
for a given fixed eccentricity and argument of pericenter,
the optimal time instances for RV observations depend 
only on the orbital phase. In the case of transiting planets
the long-term 
variations in the orbital eccentricity $e$ and longitude of pericenter 
$\varpi$ are quite relevant. The presented strategy focuses on the precise 
measurement of the Lagrangian orbital 
elements $k=e\cos\varpi$ and $h=e\sin\varpi$ (including the 
cases when the cicrular property is intended to be confirmed
at high significance).

In Section~\ref{sec:formalism}, 
the basics of the mathematical
formalism are presented, including the rules for calculating
partial derivatives. In Section~\ref{sec:miscformalism}, the
solution of the spatial problem is shown, supplemented with the inverse
problem, still using infinitely differentiable functions. The spatial
case discusses how the inclination and the argument of node should be
incorporated in the formalism without loosing the analytic property
while the inverse problem describes how the orbital elements
are derived from the coordinates and velocities.
In Section~\ref{sec:applications}, we demonstrate how this formalism
can be used for transiting extrasolar planets to efficiently 
schedule the phase of RV observations in order to minimize 
uncertainties in the Lagrangian orbital elements. 
The results are summarized in the last section.


\section{Mathematical formalism}
\label{sec:formalism}

The solution for the time evolution of Kepler's problem can be derived
in the standard way as given in various 
textbooks \citep[see, e.g.,][]{murray1999}. 
The restricted two body problem itself is an integrable ordinary differential 
equation. In the planar case, three independent integrals of motion exist
and one variable has uniform monotonicity.
The integrals are related to the well known orbital 
elements, which are used to characterize the orbit. These are
the semimajor axis $a$, the eccentricity $e$ and the longitude of 
pericenter\footnote{In two dimensions, the argument
of pericenter is always equal to the longitude of pericenter, i.e.
$\varpi\equiv\omega$} $\varpi$. 
The fourth quantity is the mean anomaly $M=nt$, where
$n=\sqrt{\mu/a^3}=2\pi/P$, the mean motion, which is zero at 
pericenter passage\footnote{Throughout this paper the mass
parameter of Kepler's problem is denoted by $\mu\equiv \mathcal{G}(m_1+m_2)$,
where $m_1$ and $m_2$ are the masses of the two orbiting bodies
and $ \mathcal{G}$ is the Newtonian gravitational constant. The
orbital period is denoted by $P$.} and $t$ is the elapsed time since
the pericenter passage.
The solution to Kepler's problem can be given in terms of the
mean anomaly $M$ as defined as
\begin{equation}
E-e\sin E=M,
\end{equation}
where $E$ is the eccentric anomaly. The planar coordinates are
\begin{eqnarray}
\xi  & = & \xi_0 \cos\varpi - \eta_0\sin\varpi, \label{ellipxi} \\
\eta & = & \xi_0 \sin\varpi + \eta_0\cos\varpi, \label{ellipeta}
\end{eqnarray}
where
\begin{eqnarray}
\xi_0  & = & a(\cos E-e), \label{ellip0xi} \\
\eta_0 & = & a\sqrt{1-e^2}\sin E \label{ellip0eta};
\end{eqnarray}
see also \citet{murray1999}, Sect.~2.4 for the derivation of these equations.
Since for circular orbits the longitude of pericenter and 
pericenter passage cannot be defined, and for nearly 
circular orbits, these can only be badly constrained; in these
cases it is useful to define a new variable, the mean 
longitude as $\lambda=M+\varpi$ to use instead of $M$. Since
$\varpi$ is an integral of the motion, $\dot\lambda=\dot M=n$. 
Therefore for circular orbits $\varpi\equiv0$ and 
\eqrefs{ellip0xi}{ellip0eta} should be replaced by
\begin{eqnarray}
\xi_0  & = & a\cos\lambda, \label{circ0xi} \\
\eta_0 & = & a\sin\lambda. \label{circ0eta}
\end{eqnarray}
To obtain an analytical solution to the problem, i.e. which is 
infinitely differentiable with respect to all of the orbital elements
and the mean longitude, first let us define the Lagrangian orbital
elements $k=e\cos\varpi$ and $h=e\sin\varpi$. 
Substituting \eqrefs{ellip0xi}{ellip0eta} into 
\eqrefs{ellipxi}{ellipeta} gives
\begin{equation}
\binom{\xi}{\eta}=a\left[\binom{c}{s}+\frac{e\sin E}{2-\ell}\binom{+h}{-k}-\binom{k}{h}\right], \label{kepler2dspatial}
\end{equation}
where $c=\cos(\lambda+e\sin E)$, $s=\sin(\lambda+e\sin E)$ and 
$\ell=1-\sqrt{1-e^2}$, the oblateness of the orbit. The derivation
of the above equation is straightforward, one should only keep in mind
that $E+\varpi=\lambda+e\sin E$. In the first part of this section
we prove that the quantities 
\begin{equation}
\eop(\lambda,k,h)=\left\{\begin{tabular}{ll} $0$ & if $k=0$ and $h=0$ \\ $e\sin E$  & otherwise\end{tabular}\right.
\end{equation}
and
\begin{equation}
\eoq(\lambda,k,h)=\left\{\begin{tabular}{ll} $0$ & if $k=0$ and $h=0$ \\ $e\cos E$  & otherwise\end{tabular}\right.
\end{equation}
are analytic -- infinitely differentiable -- 
functions of $\lambda$, $k$ and $h$ for all real values
of $\lambda$ and for all $k^2+h^2=e^2<1$. In the following parts, we utilize
the partial derivatives of these analytic functions to obtain
the orbital velocities and we derive other relations. In this section
we only deal with planar orbits, the three dimensional case is discussed
in the next section.

\subsection{Partial derivatives and the analytic property}

A real function is analytic when all of its partial derivatives
exist, the partial derivatives are continuous functions and 
only depend on other analytic functions. It is proven in 
\apref{appendixeoqeop} that the partial derivatives of 
$q=\eoq(\lambda,k,h)$ and $p=\eop(\lambda,k,h)$ are the following
for $(k,h)\ne(0,0)$:
\begin{eqnarray}
\frac{\partial q}{\partial\lambda} & = & \frac{-p}{1-q}, \label{dqdl}\\
\frac{\partial q}{\partial k} & = & \frac{c-k}{1-q} = \frac{\cos(\lambda+p)-k}{1-q}, \label{dqdk}\\
\frac{\partial q}{\partial h} & = & \frac{s-h}{1-q} = \frac{\sin(\lambda+p)-h}{1-q}  \label{dqdh}
\end{eqnarray}
and
\begin{eqnarray}
\frac{\partial p}{\partial\lambda} & = & \frac{q}{1-q}, \label{dpdl}\\
\frac{\partial p}{\partial k} & = & \frac{+s}{1-q} = \frac{+\sin(\lambda+p)}{1-q}, \label{dpdk}\\
\frac{\partial p}{\partial h} & = & \frac{-c}{1-q} = \frac{-\cos(\lambda+p)}{1-q}. \label{dpdh}
\end{eqnarray}
Since for all $k^2+h^2<1$, $q<1$ and therefore $1-q>0$, 
all of the above functions are continuous on their domains.
Since the $\sin(\cdot)$ and $\cos(\cdot)$ functions are 
analytic, therefore one can conclude that
the functions $\eoq(\cdot,\cdot,\cdot)$ and $\eop(\cdot,\cdot,\cdot)$
are also analytic. 

Substituting the definition of $p=\eop(\lambda,k,h)$ into 
\eqref{kepler2dspatial}, one can write
\begin{equation}
\binom{\xi}{\eta}=a\left[\binom{\cos(\lambda+p)}{\sin(\lambda+p)}+\frac{p}{2-\ell}\binom{+h}{-k}-\binom{k}{h}\right], \label{kepler2dspatial2}
\end{equation}
while the radial distance of the orbiting particle from the center is 
$\sqrt{\xi^2+\eta^2}=r=a(1-q)$. For small eccentricities
in \eqref{kepler2dspatial2} third term $(k,h)$ is negligible
compared to the first term $(\cos,\sin)$ while the second 
term $(h,-k)p/(2-\ell)$ is negligible compared to the third term.
Therefore for $e \ll 1$, $p$ is proportional to the difference between
the true anomaly and the mean longitude and $q$ is proportional to 
the distance offset relative to a circular orbit; both caused by 
the non-zero orbital eccentricity. 

Since \eqref{kepler2dspatial2} is a combination of purely
analytic functions, the solution of Kepler's problem is analytic
with respect to the orbital elements $a$, $(k,h)$, and to
the mean longitude $\lambda$ in the domain $a>0$ and $k^2+h^2<1$.
We note here that this formalism omits the parabolic or hyperbolic solutions.
The formalism based on the Stumpff functions \citep[see][]{stiefel1971}
provides a continuous set of formulae for the elliptic, parabolic,
and hyperbolic orbits but this parametrization is still singular 
in the $e\to0$ limit.

\subsection{Orbital velocities}

Assuming a non-perturbed orbit, i.e. when $(\dot k,\dot h)=0$, and $\dot a=0$
and when the mean motion $n=\dot\lambda$ is constant, the orbital velocities
can be directly obtained by calculating the partial derivative 
of \eqref{kepler2dspatial2} with respect to $\lambda$ and applying the 
chain rule since
\begin{equation}
\frac{\partial}{\partial t}\binom{\xi}{\eta}\equiv
\binom{\dot\xi}{\dot\eta}=
\left[\frac{\partial}{\partial \lambda}\binom{\xi}{\eta}\right]\frac{\partial\lambda}{\partial t}
=n\frac{\partial}{\partial \lambda}\binom{\xi}{\eta}. 
\end{equation}
Substituting the partial derivative \eqref{dpdl} into the
expansion of $\partial\xi/\partial\lambda$ and $\partial\eta/\partial\lambda$
one gets
\begin{equation}
\binom{\dot\xi}{\dot\eta}=\frac{an}{1-q}\left[\binom{-\sin(\lambda+p)}{+\cos(\lambda+p)}+\frac{q}{2-\ell}\binom{+h}{-k}\right]. \label{kepler2dvelocity}
\end{equation}
Note that \eqref{kepler2dvelocity} is also a combination of purely
analytic functions, the components of the orbital velocity are analytic
with respect to the orbital elements $a$, $(k,h)$, and to
the mean longitude $\lambda$.

It is also evident that the time derivative of \eqref{kepler2dvelocity} is
\begin{eqnarray}
\binom{\ddot\xi}{\ddot\eta} & = &\frac{-an^2}{(1-q)^3}\left[\binom{\cos(\lambda+p)}{\sin(\lambda+p)}+\right.  \label{kepler2daccel2} \\
& & \left.+\frac{p}{2-\ell}\binom{+h}{-k}-\binom{k}{h}\right]. \nonumber
\end{eqnarray}
Obviously, \eqref{kepler2daccel2} can be written as 
\begin{equation}
\binom{\ddot\xi}{\ddot\eta}=-\frac{n^2}{(1-q)^3}\binom{\xi}{\eta},
\end{equation}
which is equivalent to the equations of motion since 
$\mu=n^2a^3$ and $\sqrt{\xi^2+\eta^2}=r=a(1-q)$.

\subsection{Other properties}

In this subsection we summarize some other properties of the
functions $p\equiv\eop(\lambda,k,h)$ and $q\equiv\eoq(\lambda,k,h)$ 
which can also be helpful in some derivations or during numerical
evaluation.

One of the most important properties is the rotational invariance. This
is a direct consequence of the relation $M=\lambda-\varpi$, i.e. 
$q$ and $p$ would not change if the mean longitude is increased
by an arbitrary angle of $\Omega$ and simultaneously the vector $(k,h)$
is rotated with the same angle. Therefore,
\begin{eqnarray}
q & = & \eoq(\lambda-\Omega,k\cos\Omega+h\sin\Omega,-k\sin\Omega+h\cos\Omega), \label{rotq}\\
p & = & \eop(\lambda-\Omega,k\cos\Omega+h\sin\Omega,-k\sin\Omega+h\cos\Omega). \label{rotp}
\end{eqnarray}
This property also results that 
\begin{eqnarray}
q & = & \eoq(0,k\cos\lambda+h\sin\lambda,-k\sin\lambda+h\cos\lambda), \\
p & = & \eop(0,k\cos\lambda+h\sin\lambda,-k\sin\lambda+h\cos\lambda).
\end{eqnarray}

The terms $c\equiv\cos(\lambda+p)$ and $s\equiv\sin(\lambda+p)$ appear
frequently in the expressions for both the coordinates and the velocities. 
These are related to the $(q,p)$ functions and 
the orbital elements $(k,h)$ as
\begin{eqnarray}
k & = & qc+ps, \\
h & = & qs-pc
\end{eqnarray}
or similarly
\begin{eqnarray}
q & = & kc+hs, \\
p & = & ks-hc.
\end{eqnarray}
The partial derivatives of $c$ and $s$ with respect to the mean
longitude and the Lagrangian orbital elements $k$ and $h$ are
\begin{equation}
\frac{\partial}{\partial\lambda}\binom{c}{s}=\frac{1}{1-q}\binom{-s}{+c}\label{dcsdl}
\end{equation}
and
\begin{equation}
\frac{\partial(c,s)}{\partial(k,h)}=-\frac{1}{1-q}\binom{~~s^2 ~~~ -sc}{-sc ~~~ ~~c^2}.\label{dcsdkh}
\end{equation}


\section{Three dimensional case and the inverse problem}
\label{sec:miscformalism}

For an eccentric and inclined orbit, the spatial coordinates 
of an orbiting body can be obtained in
the similar manner as it is done in \citet{murray1999}, Sect.~2.8. Without
going into the details, here we present the results of the basic
calculations. The spatial coordinates $\mathbf{r}=(x,y,z)$ are
\begin{equation}
\mathbf{r}=\mathbf{P}_3\mathbf{P}_2\mathbf{P}_1\mathbf{r}_0, \label{basicspatial}
\end{equation}
where $\mathbf{r}_0=(\xi_0,\eta_0,0)$ (see also equations~\ref{ellip0xi}
and~\ref{ellip0eta}), and $\mathbf{P}_1$, $\mathbf{P}_2$ and $\mathbf{P}_3$
are the rotational matrices with respect to the argument of 
pericenter, $\omega$, the inclination, $i$, and the argument of 
the ascending node, $\Omega$. By substituting the equations for 
$(\xi_0,\eta_0)$ into \eqref{basicspatial} and using the rotational
invariance of $\eoq(\cdot,\cdot,\cdot)$ and
$\eop(\cdot,\cdot,\cdot)$, i.e. \eqrefs{rotq}{rotp}, one can derive
the spatial coordinates of the orbiting body in a similar manner as
it was done in the planar case. The result can be written in a compact
form using some other auxiliary quantities. 
First, define the Lagrangian orbital elements $i_x=2\sin(i/2)\cos\Omega$,
$i_y=2\sin(i/2)\sin\Omega$, $i_z=\sqrt{4-i_x^2-i_y^2}$, and the quantity
$W=\eta i_x-\xi i_y$, where $(\xi,\eta)$ is defined in \eqref{kepler2dspatial}.
The spatial coordinates are then 
\begin{eqnarray}
x & = & \xi +\frac{1}{2}i_yW, \label{spatix} \\
y & = & \eta-\frac{1}{2}i_xW, \label{spatiy} \\
z & = & \frac{1}{2}i_zW. \label{spatiz}
\end{eqnarray}
Since $W$ is a linear combination of $\xi$ and $\eta$,
the orbital velocities are linear combinations of 
$\dot{\xi}$ and $\dot{\eta}$, with the same coefficients, i.e.
\begin{eqnarray}
\dot x & = & \dot{\xi }+\frac{1}{2}i_y\dot W, \label{spativx} \\
\dot y & = & \dot{\eta}-\frac{1}{2}i_x\dot W, \label{spativy} \\
\dot z & = & \frac{1}{2}i_z\dot W, \label{spativz}
\end{eqnarray}
where $\dot W=\dot{\eta}i_x-\dot{\xi}i_y$.

\subsection{The inverse problem}

To compute the orbital elements $(a,\lambda,k,h,i_x,i_y)$ from 
the spatial coordinates $(x,y,z)$ and velocities $(\dot x,\dot y,\dot z)$,
first define 
\begin{eqnarray}
r^2 & = & x^2+y^2+z^2, \\
v^2 & = & \dot x^2+\dot y^2+\dot z^2, \\
c_x & = & y\dot z-z\dot y, \\
c_y & = & z\dot x-x\dot z, \\
c_z & = & x\dot y-y\dot x, \\
C^2 & = & c_x^2+c_y^2+c_z^2,  \\
\hat c & = & x\dot x+y\dot y+z\dot z.
\end{eqnarray}
The Lagrangian orbital elements for the inclination and the argument
of the ascending node is then
\begin{eqnarray}
i_x & = & -\frac{\sqrt{2}}{\sqrt{1+c_z/C}}\frac{c_y}{C}, \\
i_y & = & +\frac{\sqrt{2}}{\sqrt{1+c_z/C}}\frac{c_x}{C}.
\end{eqnarray}
For the eccentricity and the longitude of pericenter one gets
\begin{eqnarray}
\binom{k}{h}=\frac{C}{\mu}\binom{+\dot y-\frac{\dot z}{C+c_z}c_y}{-\dot x+\frac{\dot z}{C+c_z}c_x}
-\frac{1}{r}\binom{x-\frac{z}{C+c_z}c_x}{y-\frac{z}{C+c_z}c_y}. \label{khgeneral}
\end{eqnarray}
The semimajor axis satisfies the well known relation \citep{murray1999}
\begin{equation}
a=\frac{C^2}{\mu(1-k^2-h^2)}=\left(\frac{2}{r}-\frac{v^2}{\mu}\right)^{-1}.
\end{equation}
The mean longitude is then
\begin{eqnarray}
\lambda & = & \arg\left(
	r\dot y-\frac{r\dot zc_y}{C+c_z}+\frac{h\hat c}{2-\ell}\right., \nonumber \\
& & 	\left.-r\dot x+\frac{r\dot zc_x}{C+c_z}-\frac{k\hat c}{2-\ell}\right)-\frac{\hat c}{C}(1-\ell), \label{lambdageneral}
\end{eqnarray}
where $\ell=1-\sqrt{1-k^2-h^2}$. 
It can be shown that in the planar case, i.e. when $z=0$ and $\dot z=0$,
\eqref{khgeneral} and \eqref{lambdageneral} do reduce to
\begin{equation}
\binom{k}{h}=\frac{C}{\mu}\binom{+\dot y}{-\dot x}-\frac{1}{r}\binom{x}{y} \label{khplanar}
\end{equation}
and 
\begin{equation}
\lambda  =  \arg\left(r\dot y+\frac{h\hat c}{2-\ell},-r\dot x-\frac{k\hat c}{2-\ell}\right)-\frac{\hat c}{C}(1-\ell),
\end{equation}
respectively. Here, $\arg(\cdot,\cdot)$ is defined 
as $\arg(x,y)=\arctan(y/x)$ if $x\geq 0$ and $\pi+\arctan(y/x)$ otherwise.


\begin{figure}
\resizebox{8cm}{!}{\includegraphics{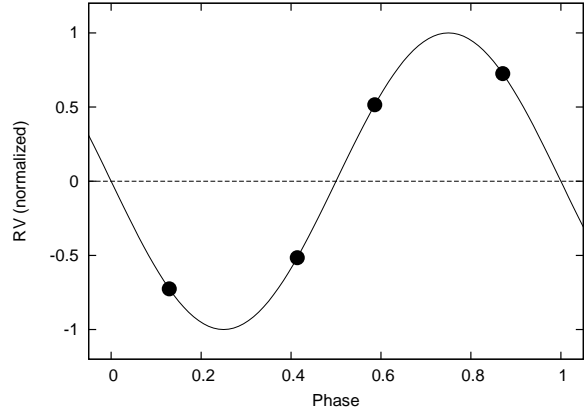}}
\caption{%
Radial velocity variations for a transiting planet orbiting its
star on a circular orbit. The transit occurs at zero (or unity) phase.
The four dots represent the phases $\varphi=0.1292$, $0.4138$, $0.5862$
and $0.8708$ when RV measurements should be 
obtained to achieve the largest significance of the orbital circularity.
}\label{fig:optphase}
\end{figure}

\section{Applications}
\label{sec:applications}

The utilization of the presented formalism can cover various aspects
of RV curve analysis. Due to the conventions and the definitions of the
orbital elements used in the astrophysics of stellar binaries,
the evaluation of \eqref{kepler2dvelocity}
simply yields the actual radial velocity as the $\dot\eta$ component
\footnote{In this case 
$an=K/\sqrt{1-e^2}$, where $K$ is the semi-amplitude of the RV curve}. 
The partial derivatives in equations (\ref{dqdl})-(\ref{dpdh}) and
in equations (\ref{dcsdl})-(\ref{dcsdkh}) can easily be used to 
calculate the parametric derivatives of the RV curves and therefore
to support various fitting algorithms 
\citep[e.g. Levenberg-Marquardt, see][]{press1992}. Moreover,
the secular variations in the radial velocities due to 
the secular changes in the orbital elements can also be 
estimated in an analytic way.

In the follow-up observations of planets discovered by transits in photometric
data series, the detection of systematic variations in the RV signal is 
one of the most relevant steps, either to rule out transits of late-type dwarf 
stars, and/or blends, or to characterize the mass of the planet and the 
orbital parameters. Since transit timing constrains the epoch and orbital
period much more precisely than radial velocity alone, 
these two can be assumed  to be fixed in the analysis of the RV data.
However, this constraint also 
includes an additional feature. 
Using equation (1) in \citet{pal2008}, 
the mean longitude at transits can be calculated as
\begin{equation}
\lambda_{\rm tr}=\arg\left(k+\frac{kh}{2-\ell},1+h-\frac{k^2}{2-\ell}\right)-\frac{k(1-\ell)}{h},
\end{equation}
therefore the mean longitude at the orbital phase $\varphi$ becomes
$\lambda=\lambda_{\rm tr}+2\pi\varphi$. Consequently,
the partial derivatives of the $\dot\eta$ RV component 
$v=\dot\eta(\lambda_{\rm tr}+2\pi\varphi,k,h)$ with respect
to the orbital elements $k$ and $h$ are
\begin{eqnarray}
\frac{\partial v}{\partial k} & = & \frac{\partial \dot\eta}{\partial k}+
	\frac{\partial \dot\eta}{\partial\lambda}\frac{\partial \lambda_{\rm tr}}{\partial k}, \\
\frac{\partial v}{\partial h} & = & \frac{\partial \dot\eta}{\partial h}+
	\frac{\partial \dot\eta}{\partial\lambda}\frac{\partial \lambda_{\rm tr}}{\partial h}.
\end{eqnarray}

A radial velocity curve of a star, caused by the perturbation of
a single companion can be parametrized by six quantities: the semi-amplitude
of RV variations, $K$, the zero point, $G$, the Lagrangian orbital
elements, $(k,h)$, the epoch, $T_0$ (or equivalently the phase at an 
arbitrary fixed time instant) and the period $P$. In the cases of 
transiting planets, the later two are known since
the transit photometric observations constrain both with exceeding 
precision (relative to the precision attainable purely by the RV data).
Therefore, one has to fit only four quantities, i.e. 
$\mathbf{a}=(K,G,k,h)$. 
The vast majority of the known transiting planets orbits their host
stars on a tight orbit and these tight orbits are expected to 
be circular (however, there are few known exceptions\footnote{see e.g. 
\texttt{http://www.exoplanet.eu} for up to date information}).

Now, we show how the optimal phases of observations can be determined
to confirm the orbital circularity, i.e.
result the smallest uncertainty in the orbital elements $k$ and $h$. 
To obtain the uncertainty of a fitted parameter and/or the
correlation between the parameters, the 
Fisher matrix method can be utilized \citep{finn1992,baluev2008}. 
This method gives the covariance matrix as 
\begin{equation}
\left<\delta a_m \delta a_n\right>=\left(\Gamma^{-1}\right)_{mn}, \label{eq:covfisher}
\end{equation}
where
\begin{equation}
\Gamma_{mn}=\sum_i \frac{\partial_m f(\mathbf{a},t_i)\partial_n f(\mathbf{a},t_i)}{\sigma_i^2}. \label{eq:fisherdef}
\end{equation}
Here $f(\mathbf{a},t)$ is the model function which depends on its
adjusted parameters $\mathbf{a}\equiv(a_1,\dots,a_N)$, and $t$ represents
the independent variable(s) (in the case of time series, there is
one independent variable, the time itself). 
Since in our case there are four unknowns, one has 
to have at least four data points, in order to completely determine the
parameters. Nevertheless, \eqref{eq:fisherdef}
can be formally evaluated providing both the uncertainties
and correlations. To determine the best four phases ($\varphi_1$,
$\varphi_2$, $\varphi_3$ and $\varphi_4$) which minimizes
the uncertainty in the eccentricity, one has to minimize 
the volume of the covariance ellipsoid of the $(k,h)$ parameters
\citep[a.k.a. generalized {\it D}-optimality, see][]{baluev2008}.
It can be shown that this volume is 
\begin{eqnarray}
U& = &\left[\det\left(
	\begin{tabular}{cc}
	$\left<\delta k^2\right>$ & $\left<\delta k\delta h\right>$  \\
	$\left<\delta h\delta k\right>$ & $\left<\delta h^2\right>$  
	\end{tabular}
\right)\right]^{1/2} = \\
& = & \sqrt{\left<\delta k^2\right>\left<\delta h^2\right>
-\left<\delta k\delta h\right>^2} \nonumber
\end{eqnarray}
where the covariances can be calculated using \eqref{eq:covfisher},
i.e. $\left<\delta k^2\right>=(\Gamma^{-1})_{33}$,
$\left<\delta k\delta h\right>=(\Gamma^{-1})_{34}$,
and $\left<\delta h^2\right>=(\Gamma^{-1})_{44}$ if 
the parameters are $\mathbf{a}=(a_1,a_2,a_3,a_4)=(K,G,k,h)$.

\begin{table}
\caption{Optimal phases of radial velocity measurements
in order to obtain the orbital eccentricity as precise
as possible. The eccentricity and the alignment of the
orbit are quantified by the Lagrangian orbital elements
$k$ and $h$.}
\label{table:optphases}
\begin{center}\begin{tabular}{rrrrrr}
\hline
\hline
$k$ & $h$ & $\varphi_1$ & $\varphi_2$ & $\varphi_3$ & $\varphi_4$ \\
\hline
$-0.4 $&$-0.4$&$ 0.1305$&$ 0.2064$&$ 0.2519$&$ 0.6943$ \\
$-0.4 $&$-0.2$&$ 0.1060$&$ 0.2048$&$ 0.2847$&$ 0.7985$ \\
$-0.4 $&$ 0.0$&$ 0.0787$&$ 0.1879$&$ 0.3125$&$ 0.8695$ \\
$-0.4 $&$ 0.2$&$ 0.0533$&$ 0.1584$&$ 0.3398$&$ 0.9197$ \\
$-0.4 $&$ 0.4$&$ 0.0316$&$ 0.1180$&$ 0.3701$&$ 0.9555$ \\
$-0.2 $&$-0.4$&$ 0.1964$&$ 0.3307$&$ 0.3910$&$ 0.7027$ \\
$-0.2 $&$-0.2$&$ 0.1497$&$ 0.3180$&$ 0.4207$&$ 0.7943$ \\
$-0.2 $&$ 0.0$&$ 0.1076$&$ 0.2927$&$ 0.4522$&$ 0.8616$ \\
$-0.2 $&$ 0.2$&$ 0.0722$&$ 0.2551$&$ 0.4900$&$ 0.9113$ \\
$-0.2 $&$ 0.4$&$ 0.0437$&$ 0.2040$&$ 0.5399$&$ 0.9481$ \\
$ 0.0 $&$-0.4$&$ 0.2557$&$ 0.4672$&$ 0.5328$&$ 0.7443$ \\
$ 0.0 $&$-0.2$&$ 0.1854$&$ 0.4445$&$ 0.5555$&$ 0.8146$ \\
$ 0.0 $&$ 0.0$&$ 0.1292$&$ 0.4138$&$ 0.5862$&$ 0.8708$ \\
$ 0.0 $&$ 0.2$&$ 0.0850$&$ 0.3728$&$ 0.6272$&$ 0.9150$ \\
$ 0.0 $&$ 0.4$&$ 0.0511$&$ 0.3169$&$ 0.6831$&$ 0.9489$ \\
$ 0.2 $&$-0.4$&$ 0.2973$&$ 0.6090$&$ 0.6693$&$ 0.8036$ \\
$ 0.2 $&$-0.2$&$ 0.2057$&$ 0.5793$&$ 0.6820$&$ 0.8503$ \\
$ 0.2 $&$ 0.0$&$ 0.1384$&$ 0.5478$&$ 0.7073$&$ 0.8924$ \\
$ 0.2 $&$ 0.2$&$ 0.0886$&$ 0.5100$&$ 0.7449$&$ 0.9278$ \\
$ 0.2 $&$ 0.4$&$ 0.0519$&$ 0.4601$&$ 0.7960$&$ 0.9563$ \\
$ 0.4 $&$-0.4$&$ 0.3057$&$ 0.7481$&$ 0.7936$&$ 0.8695$ \\
$ 0.4 $&$-0.2$&$ 0.2016$&$ 0.7153$&$ 0.7952$&$ 0.8939$ \\
$ 0.4 $&$ 0.0$&$ 0.1305$&$ 0.6875$&$ 0.8121$&$ 0.9212$ \\
$ 0.4 $&$ 0.2$&$ 0.0803$&$ 0.6602$&$ 0.8416$&$ 0.9467$ \\
$ 0.4 $&$ 0.4$&$ 0.0445$&$ 0.6299$&$ 0.8820$&$ 0.9684$ \\
\hline
\hline
\end{tabular}\end{center}
\end{table}

We minimized $U$ as the function of the phases $\varphi_1$, \dots,
$\varphi_4$ using the Markov chain Monte-Carlo (MCMC) method 
\citep[see e.g.][]{ford2004}. Multiple chains were initiated 
from four random phases (namely, all of them are chosen uniformly from
the interval $[0,1]$). We found that all of the chains converge to
a single set of phases that represent the minimum of the
function $U(\varphi_1,\varphi_2,\varphi_3,\varphi_4)$. 
Therefore, the set of optimal phases is
unique: we found the optimal phases are $\varphi=0.1292$, $0.4138$, $0.5862$ 
and $0.8708$. In Fig.~\ref{fig:optphase} these phases are marked on 
a hypothetical radial velocity curve. We note that this volume of
the covariance ellipsoid of the $(k,h)$ parameters is approximately
$12$ times smaller on the average if the phases were chosen randomly,
and $2.5$ times smaller if the phases were chosen to be closer
with a factor of $2$ to the phases $0.25$ and $0.75$ 
(i.e. $\varphi'=0.1896$, $0.3319$ and $0.6681$, $0.8104$, respectively). 

Of course, the same kind of calculation of the phases that yields 
the smallest combined uncertainty in the $(k,h)$ parameters 
can be performed for arbitrary orbital eccentricity. 
For some certain values of $(k,h)$, 
these optimal phases are shown in Table~\ref{table:optphases}. 
Thus, if some initial values for the orbital elements $k$ and $h$
are known, further observations can be planned accordingly.

Similarly, the optimal phases can be derived for arbitrary number 
of observations ($N_{\rm obs}$). However, it turns out that
for $N_{\rm obs}\ge 5$, the phase volume $U$ have more than one local 
minima. In order to find the global minimum, we initiated several 
hundreds or thousands of individual initial conditions and applied 
both the previously discussed MCMC
method and the downhill simplex algorithm \citep{press1992} in order
to find the global minimum of $U$. For $N_{\rm obs}=5$, there are 
two local minima and the global minimum is at the 
phases $\varphi=(0.1318, 0.3978, 0.5, 0.6022, 0.8682)$. For $N_{\rm obs}>5$,
one or more of the phases will be degenerated. In other words, 
one should take RV measurements at the same phases in order to 
minimize the uncertainty in the orbital eccentricity. For instance,
the global minimum is at
$\varphi=(0.1376, 0.4204, 0.4204, 0.5796, 0.5796, 0.8624)$ for
$N_{\rm obs}=6$ or
$\varphi=(0.1405, 0.4315, 0.4315, 0.5965, 0.5965, 0.8746, 0.8746)$
for $N_{\rm obs}=7$. It can be shown that for larger $N_{\rm obs}$
which are multiple of $4$, the optimal phases will be at the exactly
same location as in the case of $N_{\rm obs}=4$ (see earlier 
or Table~\ref{table:optphases}), and at each phase one should 
acquire the same number of measurements, namely $N_{\rm obs}/4$.

Due to the limitations in the telescope time and in the 
observation conditions (including day/night variations or the visibility
of the target object), RV observations cannot be scheduled at the
optimal phases. In practice, we have a series of RV data points
and then the upcoming measurements are intended to be acquired
to yield the smallest uncertainty in the orbital eccentricity 
(or any other parameters that are the points of interest). Since
this set of phases yields enormous amount of free parameters on 
which the phases of the upcoming measurements are depend,
these phases have to derived independently by hand for each case.
The \texttt{optsn} code\footnote{http://szofi.elte.hu/\~{ }apal/utils/astro/eof/}
is intended to derive these optimal 
phases\footnote{This code was also used 
to derive the previously discussed orbital 
phases as well as the values found in Table~\ref{table:optphases}.} 
for arbitrary set of fixed observations.

Since in a multiple planetary system,
secular perturbations result the highest variations in the orbital elements
$(k,h)$, the proper selection of RV measurement instants can significantly
increase the detection probability of other companions. 


\section{Summary}
\label{sec:discussion}

Although the Newtonian gravitational two-body 
problem (a.k.a.,~Kepler's problem)
is integrable, its solution requires transcendent equations.
Moreover the parameter of these equations has a finite discontinuity in
the limit of circular orbits, and this ill-behaved property leads to
numerical disadvantages. For instance the calculation of the optimal
phases, as presented in Section~\ref{sec:applications}, cannot be 
performed if the orbit was parametrized by $(e,\varpi)$ instead of 
$(k,h)$ since the partial derivative $\partial v/\partial e|_{e=0}$,
which is required to be known to obtain the Fisher-matrix, does not exist. 

In this paper a new framework has been presented to describe 
the time evolution of Kepler's problem
using analytic functions which omits singular parametrization. This formalism
is then used to construct orbital solution for three dimensional orbits
in a similar manner. Using this analytic solution, it is straightforward
to derive the radial velocity function, which is one of the most relevant
observable quantities in the physics of stellar binaries and extrasolar planets.
The functions $\eop(.,.,.)$ and $\eoq(.,.,.)$ can be handled as an 
analytic function, just as simply as if they were a trigonometric 
function in practice. It is also straightforward
to carry out a Fisher analysis of RV data to estimate the expected uncertainty
of the physical parameters. The implementation of the
functions $\eop(.,.,.)$ and $\eoq(.,.,.)$ are simple in any kind
of programming languages, a demonstration code 
is provided in the \texttt{gnuplot} 
language\footnote{http://szofi.elte.hu/\~{ }apal/utils/astro/eof/eof.gnuplot}.


\section*{Acknowledgments}

The author thanks Dan Fabrycky and Bence Kocsis for the useful
discussions, and also thanks G\'asp\'ar Bakos
for advices about practical implementation.
The author would like to thank the referee for the suggestions
about the improvements related to the possible applications.
The author acknowledges the support by the HATNet project, 
and NASA grant NNG04GN74G. This work was also supported by
ESA grant PECS~98073. 


{}


\onecolumn
\appendix

\section{The partial derivatives of the eccentric offset functions}
\label{appendixeoqeop}

We know that the eccentric offsets $q$ and $p$ are continuous functions of
$\lambda$, $k$ and $h$. Now the  partial derivatives
of $q\equiv\eoq(\lambda,k,h)$ and $p\equiv\eop(\lambda,k,h)$ are calculated. 
If all of the partial derivatives are 
non-singular around zero (where $k=h=0$), one could conclude that the 
eccentric offsets are not only continuous but smooth functions. 
For the derivation of the partial derivatives 
\begin{equation}
\frac{\partial(q,p)}{\partial(\lambda,k,h)}\equiv
\left(\begin{tabular}{ccc}
        $\displaystyle\frac{\partial q}{\partial \lambda}$ &
	$\displaystyle\frac{\partial q}{\partial k}$ & 
        $\displaystyle\frac{\partial q}{\partial h}$ \\[4mm]
	$\displaystyle\frac{\partial p}{\partial \lambda}$ & 
        $\displaystyle\frac{\partial p}{\partial k}$ & 
        $\displaystyle\frac{\partial p}{\partial h}$ 
\end{tabular}\right)
\end{equation}
we will use implicit function theorem.
Let us define
\begin{eqnarray}
F_x(q,p;\lambda,k,h) & = & +q\cos p+p\sin p - (k\cos\lambda+h\sin\lambda), \\
F_y(q,p;\lambda,k,h) & = & -q\sin p+p\cos p - (k\sin\lambda-h\cos\lambda).
\end{eqnarray}
For a fixed value of $\lambda$, $k$ and $h$, Kepler's equation is 
equivalent with
\begin{equation}
\binom{F_x(q,p;\lambda,k,h)}{F_y(q,p;\lambda,k,h)}=0.
\end{equation}
According to the implicit function theorem, the required partial derivatives
are 
\begin{equation}
\left(\begin{tabular}{ccc}
	$\displaystyle\frac{\partial q}{\partial \lambda}$ &
	$\displaystyle\frac{\partial q}{\partial k}$ & 
	$\displaystyle\frac{\partial q}{\partial h}$ \\[4mm]
	$\displaystyle\frac{\partial p}{\partial \lambda}$ &
	$\displaystyle\frac{\partial p}{\partial k}$ & 
	$\displaystyle\frac{\partial p}{\partial h}$
\end{tabular}\right)=-
\left(\begin{tabular}{cc}
	$\displaystyle\frac{\partial F_x}{\partial q}$ & 
	$\displaystyle\frac{\partial F_x}{\partial p}$ \\[4mm]
	$\displaystyle\frac{\partial F_y}{\partial q}$ & 
	$\displaystyle\frac{\partial F_y}{\partial p}$ 
\end{tabular}\right)^{-1}\times
\left(\begin{tabular}{ccc}
	$\displaystyle\frac{\partial F_x}{\partial \lambda}$ &
	$\displaystyle\frac{\partial F_x}{\partial k}$ & 
	$\displaystyle\frac{\partial F_x}{\partial h}$ \\[4mm]
	$\displaystyle\frac{\partial F_y}{\partial \lambda}$ &
	$\displaystyle\frac{\partial F_y}{\partial k}$ & 
	$\displaystyle\frac{\partial F_y}{\partial h}$ 
\end{tabular}\right). \label{eq:dpqimplicit}
\end{equation}
The partial derivatives on the right-hand side of \eqref{eq:dpqimplicit}
are
\begin{equation}
\left(\begin{tabular}{cc}
	$\displaystyle\frac{\partial F_x}{\partial q}$ & 
	$\displaystyle\frac{\partial F_x}{\partial p}$ \\[4mm]
	$\displaystyle\frac{\partial F_y}{\partial q}$ & 
	$\displaystyle\frac{\partial F_y}{\partial p}$ 
\end{tabular}\right)
=
\left(\begin{tabular}{cc}
	$\displaystyle+\cos p$ & 
	$\displaystyle-q\sin p+\sin p+p\cos p$ \\
	$\displaystyle-\sin p$ & 
	$\displaystyle+\cos p-p\sin p-q\cos p$ 
\end{tabular}\right)
\end{equation}
and
\begin{equation}
\left(\begin{tabular}{ccc}
	$\displaystyle\frac{\partial F_x}{\partial \lambda}$ &
	$\displaystyle\frac{\partial F_x}{\partial k}$ & 
	$\displaystyle\frac{\partial F_x}{\partial h}$ \\[4mm]
	$\displaystyle\frac{\partial F_y}{\partial \lambda}$ &
	$\displaystyle\frac{\partial F_y}{\partial k}$ & 
	$\displaystyle\frac{\partial F_y}{\partial h}$ 
\end{tabular}\right)
=
\left(\begin{tabular}{ccc}
	$\displaystyle+k\sin\lambda-h\cos\lambda$ &
	$\displaystyle-\cos\lambda$ &
	$\displaystyle-\sin\lambda$ \\
	$\displaystyle-k\cos\lambda-h\sin\lambda$ &
	$\displaystyle-\sin\lambda$ &
	$\displaystyle+\cos\lambda$ 
\end{tabular}\right).
\end{equation}
Therefore, expanding the above partial derivatives, one can obtain that
\begin{equation}
\left(\begin{tabular}{ccc}
	$\displaystyle\frac{\partial q}{\partial \lambda}$ &
	$\displaystyle\frac{\partial q}{\partial k}$ & 
	$\displaystyle\frac{\partial q}{\partial h}$ \\[4mm]
	$\displaystyle\frac{\partial p}{\partial \lambda}$ &
	$\displaystyle\frac{\partial p}{\partial k}$ & 
	$\displaystyle\frac{\partial p}{\partial h}$
\end{tabular}\right)=
\frac{1}{1-q}
\left(\begin{tabular}{ccc}
	$\displaystyle -p$ &
	$\displaystyle\cos(\lambda+p)-k$ &
	$\displaystyle\sin(\lambda+p)-h$ \\
	$\displaystyle+q$ &
	$\displaystyle+\sin(\lambda+p)$ &
	$\displaystyle-\cos(\lambda+p)$
\end{tabular}\right)
\end{equation}

\bsp

\label{lastpage}

\end{document}